\documentclass[cits]{PoS}

\usepackage{subfigure}
\usepackage{multirow}

\title{Aspects of Precision Calculations of Nucleon Generalized Form Factors with Domain Wall Fermions on an Asqtad Sea\footnote{Combined proceedings for talks titled "Nucleon Electromagnetic Form Factors with Domain Wall Fermions on an Asqtad Sea" and "Nucleon Generalized Form Factors with Domain Wall Fermions on an Asqtad Sea".}
}

\ShortTitle{Nucleon Generalized Form Factors}

\author{LHP Collaboration: J. D. Bratt$^a$,
R. G. Edwards$^b$,
M. Engelhardt$^c$,
G. T. Fleming$^d$,\hspace{.2cm}
Ph. H\"agler$^e$,
M. F. Lin\thanks{Speakers.}\,\,$^a$,
H. B. Meyer$^a$,
B. Musch$^e$,
{J. W. Negele}$^{\dagger a}$,
K. Orginos$^f$,\hspace{1.5 cm}
A. V. Pochinsky$^a$,
M. Procura$^a$,
D. B. Renner$^g$,
D. G. Richards$^b$,
W. Schroers$^h$,\hspace{1.5cm}
and S. Syritsyn$^a$\\
\llap{$^a$}Center\hspace{-0.0725pc} for\hspace{-0.0725pc} Theoretical\hspace{-0.0725pc} Physics,\hspace{-0.0725pc} Massachusetts\hspace{-0.0725pc} Institute\hspace{-0.0725pc} of\hspace{-0.0725pc} Technology,\hspace{-0.0725pc} Cambridge,\hspace{-0.0725pc} MA\hspace{-0.0725pc} 02139,\hspace{-0.0725pc} USA\\
\llap{$^b$}Thomas Jefferson National Accelerator Facility, Newport News, VA 23606, USA \\
\llap{$^c$}Department of Physics, New Mexico State University, Las Cruces, NM 88003-8001, USA \\
\llap{$^d$}Sloane Physics Laboratory, Yale University, New Haven, CT 06520, USA \\
\llap{$^e$}Institut f\"ur Theoretische Physik T39, TU M\"unchen, D-85747 Garching, Germany \\
\llap{$^f$}Department of Physics, College of William and Mary, Williamsburg, VA 23187, USA \\
\llap{$^g$}DESY Zeuthen, Theory Group, Platanenallee 6, D-15738 Zeuthen, Germany  \\
\llap{$^h$}Department of Physics, 
 National Taiwan University, Taipei 10617, Taiwan \\
}

\abstract{In order to advance lattice calculations of moments  of unpolarized, helicity, and transversity distributions, electromagnetic form factors, and generalized form
factors of the nucleon to a new level of precision, this work investigates several key aspects of precision lattice calculations. We calculate the number of configurations required for constant statistical errors as a function of pion mass, describe the coherent sink method to help achieve these statistics, examine the statistical correlations between separate measurements, study correlations in the behavior of form factors at different momentum transfer, examine volume dependence, and compare mixed action   results with those using comparable dynamical domain wall configurations.  We also show selected form factor results and comment on the QCD evolution of our calculations of the flavor non-singlet nucleon angular momentum.

\vspace{-20cm}\parbox{\textwidth}{\flushright\large\rm \hfill MIT-CTP 3991}\vspace{20cm}
}

\FullConference{The XXVI International Symposium on Lattice Field Theory\\
		 July 14-19 2008\\
		 Williamsburg, Virginia, USA}

\begin{document}

$\phantom{\speaker{M.F.~Lin and J.W.~Negele}}$

\section{Introduction}

Understanding the quark and gluon structure of the nucleon is a vital component of our endeavor to understand how QCD gives rise to the properties of the observed universe and is the focus of frontier experiments in nuclear and particle physics.   Lattice QCD provides a unique tool to study  nucleon structure from first principles, and as summarized in the  plenary talk at this conference by J. Zanotti \cite{Zanotti} many successful  techniques have been developed to calculate  form factors and generalized form factors of the nucleon. With the development and availability of Petascale computer resources,  we are now entering an era  in which high precision calculations of nucleon structure are becoming feasible, which necessitates a fresh examination of statistical and systematic uncertainties. Here we describe the results of several developments in our collaboration's efforts to enter a new regime of precision in calculating hadron structure. The results in this talk will focus on calculations with a hybrid action
combining domain wall valence fermions with improved staggered sea quarks, as described in  Ref.~\cite{Hagler:2007xi}, and that utilize the extensive set of configurations with dynamical improved staggered quarks generated by
the MILC collaboration~\cite{Bernard:2001av}.  We will also refer to  recent calculations using dynamical domain wall configurations \cite{Antonio:2006px,Christ:2006zz} described in the talk by S. Syritsyn \cite{Syritsyn1}.

We emphasize some technical aspects of the nucleon form factor calculation which are necessary for good controls of both statistical errors and lattice artifacts. A more thorough description of the calculation details and simulation parameters of our mixed action project can be found in Refs.~\cite{Hagler:2007xi} and \cite{WalkerLoud:2008bp}. This proceedings is organized as follows: In Section~\ref{sec:cost} we present an analysis which estimates the increase of the numerical cost as we go to the physical pion mass. In Section~\ref{sec:technique} we discuss an improved technique in the calculation of the backward propagators for the nucleon three-point correlation functions. Studies of correlations in the lattice data are given in Section~\ref{sec:stat}. We discuss possible finite volume effects and discretization errors of our calculations in Section~\ref{sec:artifacts}. We show selected form factor results and comment on the QCD evolution of our calculations of the flavor non-singlet nucleon angular momentum in Section~\ref{sec:results}, followed by conclusions in Section~\ref{sec:conclusions}.


\section{Numerical Cost of Precision Calculations}
\label{sec:cost}

As we progress to ever lighter pion masses, it is important to quantitatively estimate the
statistics required to achieve a specified level of precision as a function of pion mass. From the perspective of fluctuations produced by a set of gauge configurations, the three-point function and corresponding two-point function with the same source sink separation behave similarly, so the two-point correlation functions  $ C_2(t) =
\langle J(t) J(0)\rangle $, where $J$ denotes the nucleon source,  are a useful measure of the statistical fluctuations.  Since the variance in $C_2(t)$ is generated by a source containing three quarks and three antiquarks, it receives contributions from both three-pion and two-nucleon states, and the corresponding signal to noise ratio is therefore given by
\[
\frac{\rm Signal}{\rm Noise}  =   \frac{ \langle J(t) J(0) \rangle} { \frac{1}{\sqrt{N}}  \sqrt{ 
\langle |J(t) J(0) |^2 \rangle 
- (\langle J(t) J(0) \rangle )^2   } } 
 \sim  \frac{A e^{-M_Nt}}{ \frac{1}{\sqrt{N}}\sqrt{B e^{-3 m_\pi t } - Ce^{-2 M_N t}    }   }  \sim \sqrt{N} De^{-(M_N- \frac{3}{2} m_\pi)t} \nonumber ,
\]%
%
%
%
%
\begin{figure}[ht]
  \hfill
  \centering
\subfigure[]{
      \includegraphics[height=11pc,width=0.45\textwidth]{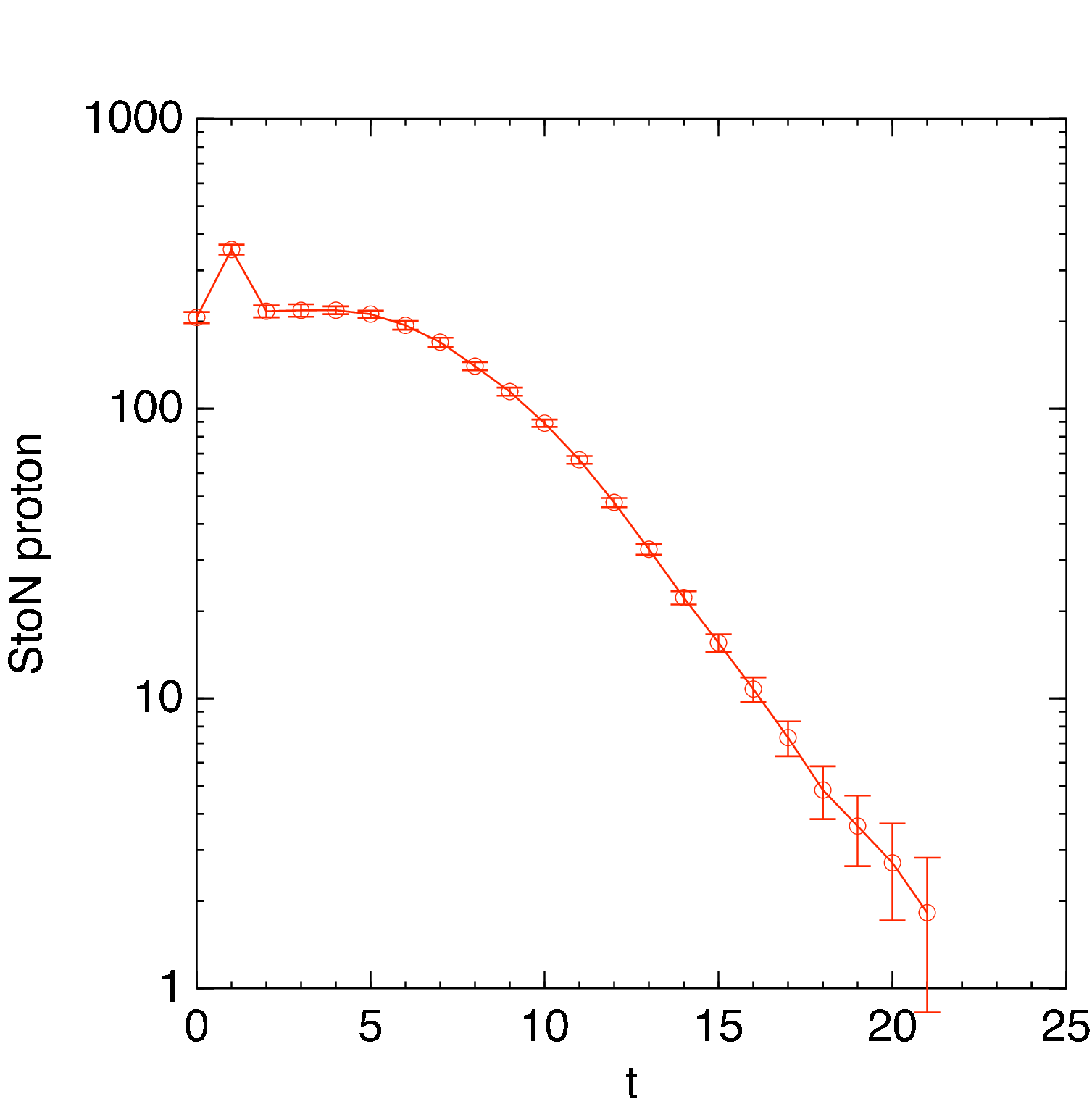}
      \label{fig:signal2noise_t}
 }
 \hfill
 \subfigure[]{
      \includegraphics[height=11pc,width=0.45\textwidth]{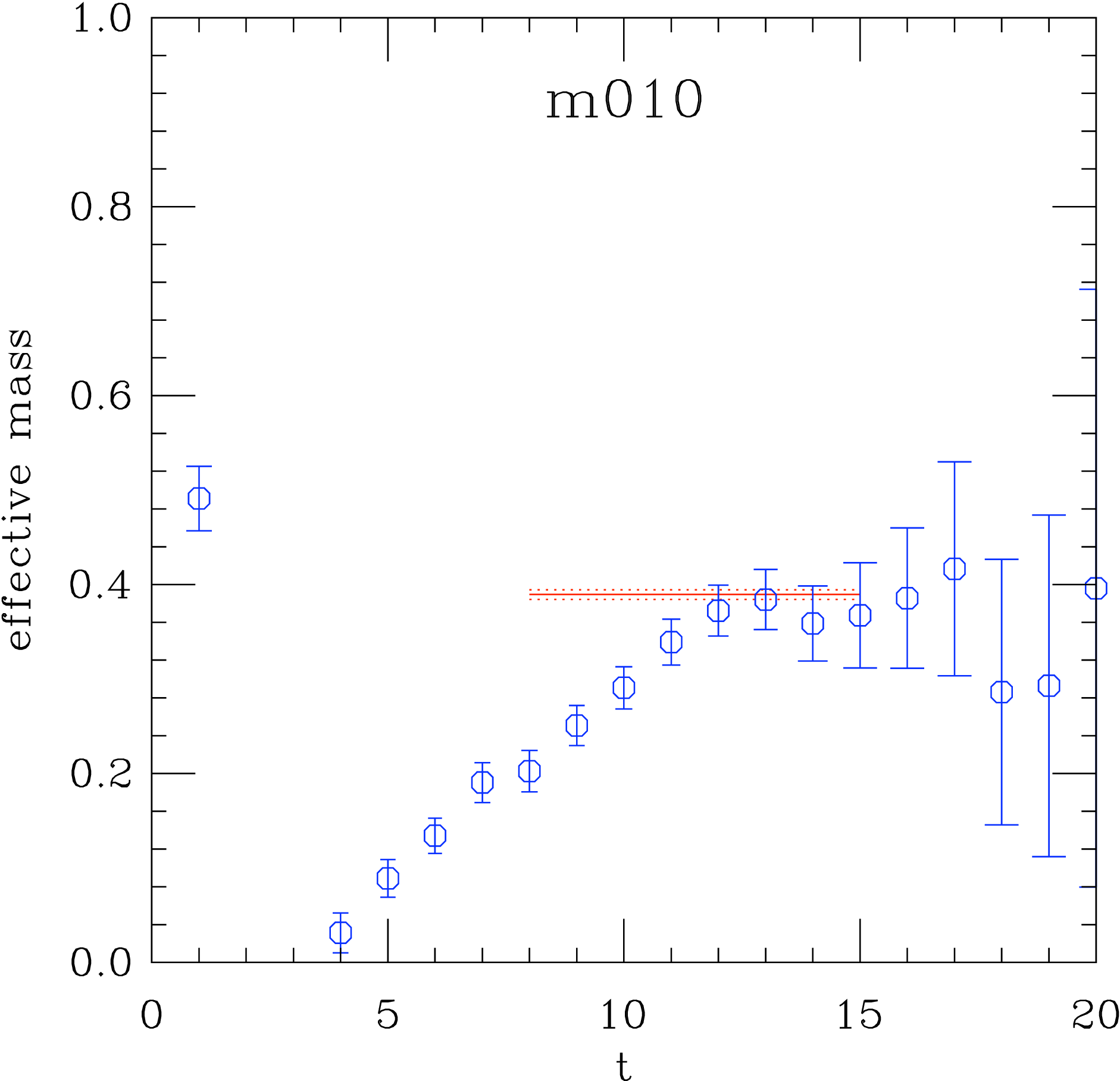}
      \label{fig:signal2noise_effm}
 }
  \caption{ The left panel shows the signal-to-noise ratio of the nucleon two-point correlation function on Asqtad lattices at $m_\pi  \sim $350 MeV, where beyond $t$ = 12, the exponential decay is given by $(M_N-\frac{3}{2} m_\pi)$. The right panel shows the corresponding ``effective mass'' of the signal-to-noise ratio in (a), where the horizontal line is the measured $M_N - \frac{3}{2} m_\pi$ of the ensemble. }
\end{figure}
\begin{figure}[ht]
\centering
\subfigure[]{
       \includegraphics[height=12pc,width=0.5\textwidth]{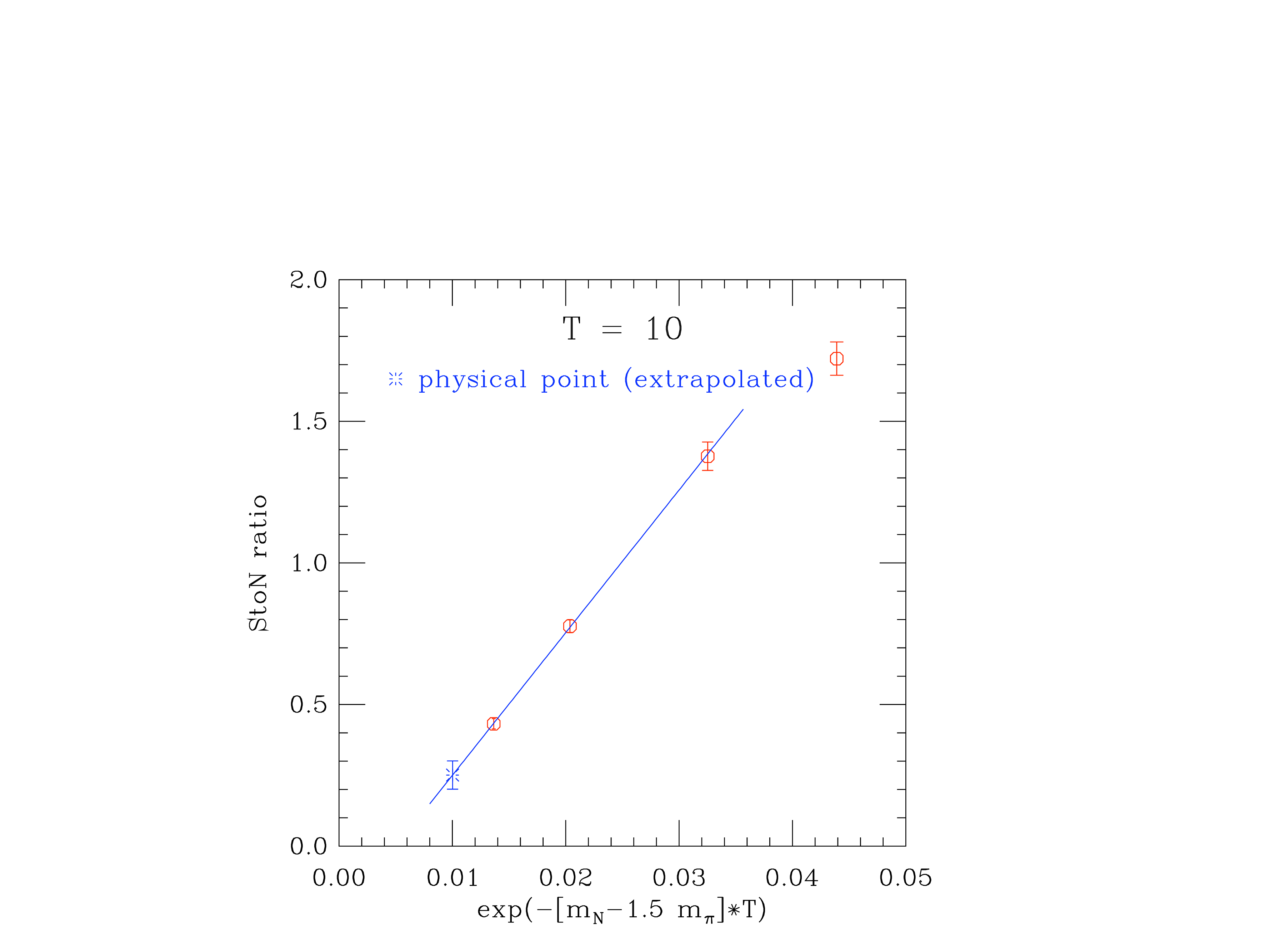}
       \label{fig:signal2noise_extrap}
  }
 \hfill
 \subfigure[]{
       \includegraphics[height=12pc,width=0.45\textwidth]{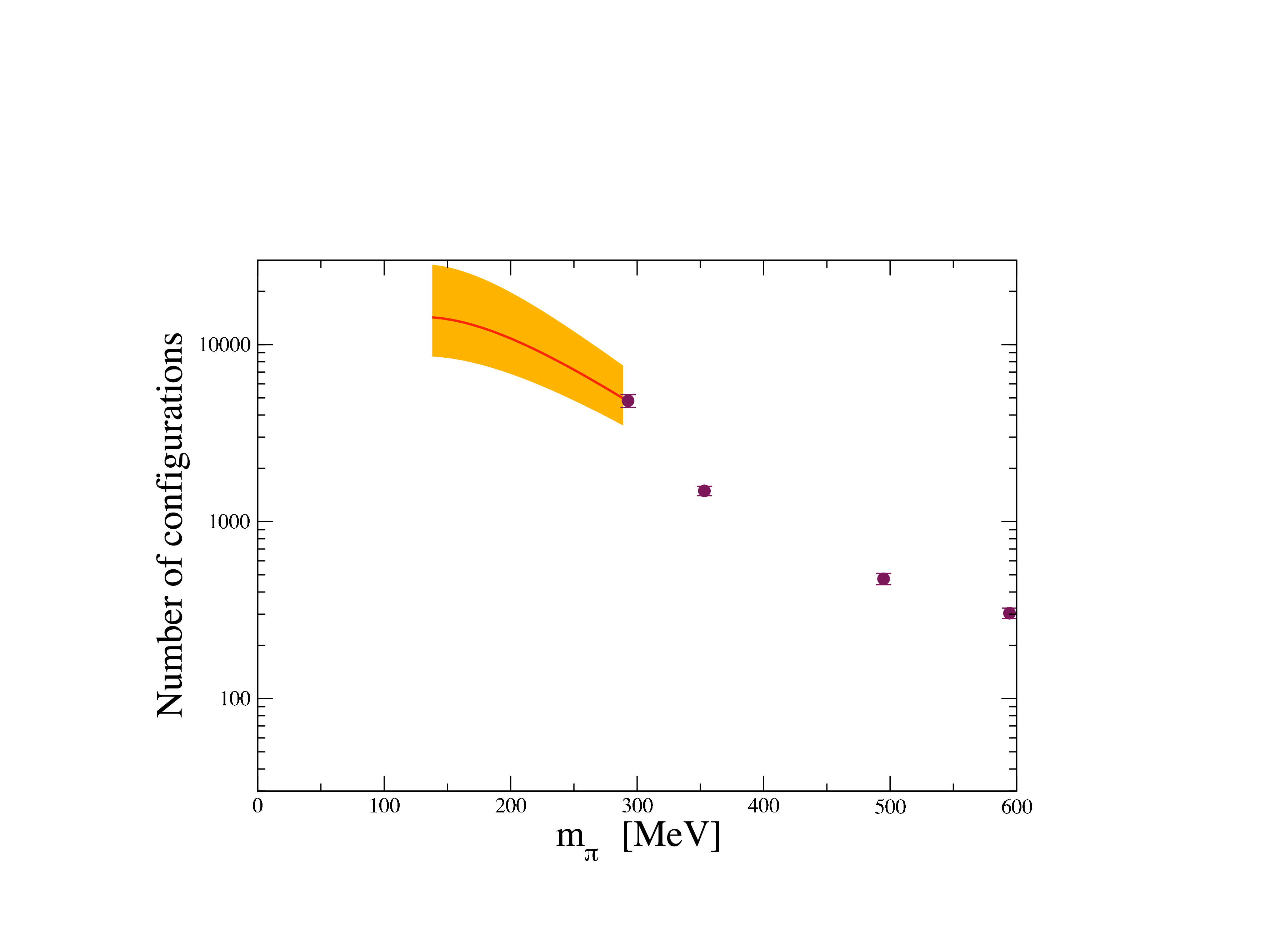}
       \label{fig:signal2noise_mpi}
  }
 \caption{The left panel shows the extrapolation of the signal-to-noise ratio to the physical pion mass. The right panel shows the exponential increase of the number of configurations needed to maintain 3\% accuracy at separation $t$= 10 as one approaches the chiral limit.}
 \end{figure}
which decreases exponentially with $(M_N-\frac{3}{2}m_\pi)t$. Figure~\ref{fig:signal2noise_t} shows a typical result for the signal to noise ratio as a function of the time separation, which displays the expected exponential decay in $M_N-\frac{3}{2} m_\pi $ at large $t$. This is clear from Figure~\ref{fig:signal2noise_effm}, where we show the ``effective mass'' of the signal to noise, compared with the measured $M_N - \frac{3}{2} m_\pi$ of the ensemble, which is denoted by the horizontal line. Using comparable calculations for the three lowest pion masses, $ \sim $ 300, 350 and 500 MeV,  we can extrapolate the signal-to-noise ratio to the physical point, as shown in Figure~\ref{fig:signal2noise_extrap}. Correspondingly, the number of configurations required to attain 3\% accuracy is shown in Figure~\ref{fig:signal2noise_mpi}.

\section{Coherent Sink Techniques to Increase Statistics}
\label{sec:technique}

Given the need for 5,000 to 10,000 independent measurements to overcome the exponentially increasing noise produced by three-pion states as the pion mass is decreased, we generate 8 independent measurements on a lattice of time extent $T= 64$ as follows.  On the first configuration, we place sources at space-time positions $ (\vec 0, 0)$, $ (\vec L/2, 16)$, $ (\vec 0, 32)$, and $ (\vec L/2, 48)$, and calculate 12 sets of propagators which we will refer to as forward propagators in the usual way. Using the forward propagators from the $i^{th}$ source $ (\vec x_i , T_i)$, we create a momentum projected nucleon sink at time  $T_0$ away  at $ (\vec x_i , T_i + T_0)$.  A conventional calculation would combine forward propagators from the source at $ (\vec x_i , T_i)$ and backward propagators from sink at $ (\vec x_i , T_i+T_0)$ to obtain the relevant two-point function, requiring 4 sets of inversions to treat all 4 sources.   In contrast, we calculate a single set of coherent backward propagators in the simultaneous presence of all 4 sources.  Combining these coherent backward propagators with the forward propagators from the $i^{th}$ source yields the physical result for the $i^{th}$ source with the $i^{th}$ sink plus terms that vanish by gauge invariance when averaged over an ensemble of configurations.  In addition, on the same lattice, 
using the forward propagators from the $i^{th}$ source $ (\vec x_i , T_i)$, we also create a momentum projected antinucleon sink a time  $T_0$ away  at $ (\vec x_i , T_i - T_0)$, 
and perform an analogous calculation for coherent antinucleon propagators. It is straightforward to relate the matrix elements of our twist-two quark operators in an antinucleon to the desired results in a nucleon.  The net result is that, given a set of forward propagators,  we obtain eight measurements at the cost of two rather than eight sets of inversions.  

To minimize correlations, the sources on the next configuration in the ensemble to be analyzed are located at space-time positions $ (\vec L/2, 0)$, $ (\vec 0, 16)$, $ (\vec L/2, 32)$, and $ (\vec 0, 48)$, and subsequent configurations are each shifted by a displacement $L/2$. The independence of these lattice measurements is addressed in the next section.
We used these coherent sink techniques on the three lowest pion masses, the parameters of which are given in Table~\ref{tab:pars}.

 \begin{table}[t]
 \centering
      \begin{tabular}{cccccc}
          \hline
          $L^3\times T$ & $(am_l)/(am_s)^{\rm asqtad}$ &  $m_\pi^{\rm DWF}$ [MeV]  & \# confs & \# meas \\
          \hline
          $20^3\times 64$ &  0.007/0.05 &293 & 464 & 3712\\
          $20^3\times 64$  & 0.01/0.05 & 356& 628  & 5024\\
          $28^3\times 64$  & 0.01/0.05 & 353& 274  & 2192\\
          \hline
      \end{tabular}
      \label{tab:pars}
      \caption{Numbers of measurements for the three lowest pion masses.}
 \end{table}

\section{Statistical Analysis}
\label{sec:stat}
\subsection{Binning and Autocorrelations}

\begin{figure}[t]
  \centering
  \includegraphics[width=0.45\textwidth,angle=-90]{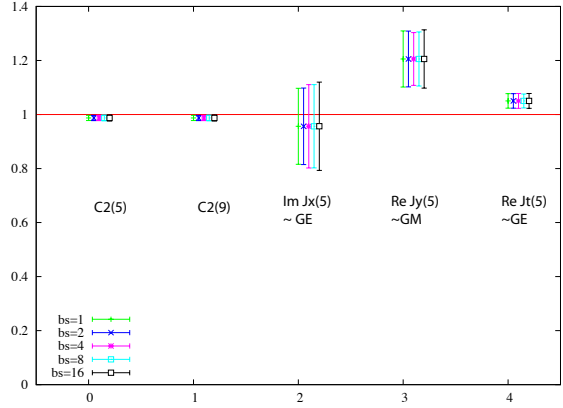}
  \caption{\label{bin}Jackknife errors with bin sizes ranging from 1 to 16 measurements as described in the text. Independence of the error with bin size indicates negligible  correlations between measurements. }
\end{figure}

A crucial question concerning our calculations with 8 measurements per lattice is the statistical independence of measurements within a single lattice and between subsequent lattices. One standard test of correlations is binning potentially correlated measurements and observing the dependence of the jackknife errors on the bin size.  Figure~\ref{bin} shows the results of  measurements with five different binnings for the two point function $C_2$ measured midway between source and sink ($t=5$) and at the source-sink separation ($t=9$), and three current operators, $J_x$, $J_y$, and  $J_t$, measured midway between the source and sink ($t=5$) measuring the electric form factor, $G_E$, the magnetic form factor, $G_M$, and  $G_E$, respectively.  Bin size 1 treats each measurement separately, size 2 combines nucleon and antinucleon from the same source, size 4 combines two nucleon and two antinucleons, size 8 include all 8 nucleons and antinucleons on a single lattice, and size 16 combines two sequential lattices. As is clear from Figure ~\ref{bin} the negligible change in the errors with bin size indicates negligible correlations.

One can also check the correlations between the measurements by calculating the integrated autocorrelation time~\cite{Wolff:2003sm}, defined as 
\begin{equation}
\tau_{\rm int} = 1/2+ \sum_{n=1}^{N_{cut}} {\rho(n)}/{\rho(0)},
\end{equation}
where $\rho(n)$ is the autocorrelation function between $n$th and $0$th measurements. The separation between two independent measurements would be $2\tau_{\rm int}$. 
\begin{figure}[t]
\centering
\subfigure[$\tau_{\rm int}$ for two-point functions.]{ 
\includegraphics[angle=-90,width=0.45\textwidth]{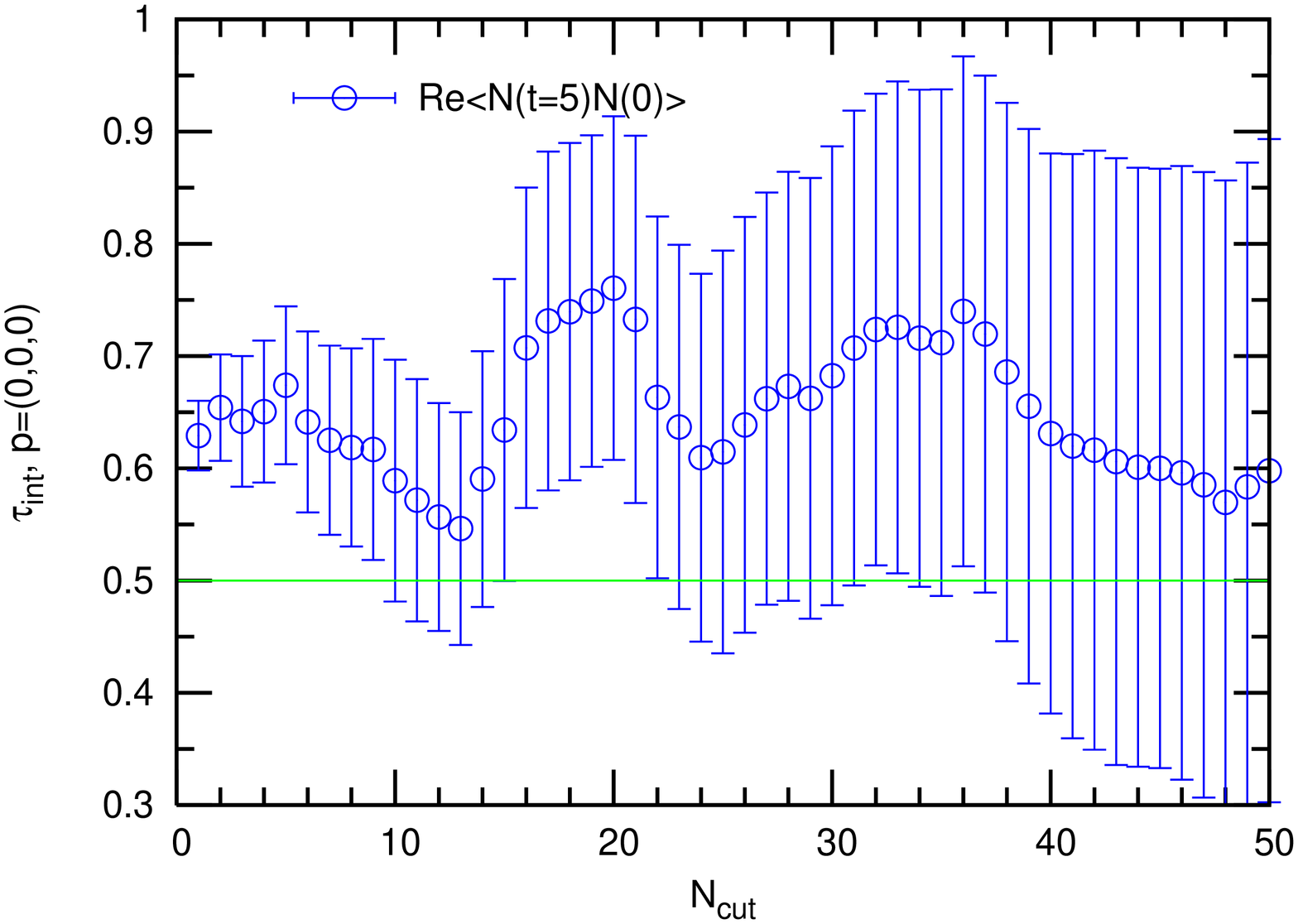}
\label{fig:int_twopt}
}
\subfigure[$\tau_{\rm int}$ for three-point functions.]{
 \includegraphics[angle=-90,width=0.45\textwidth]{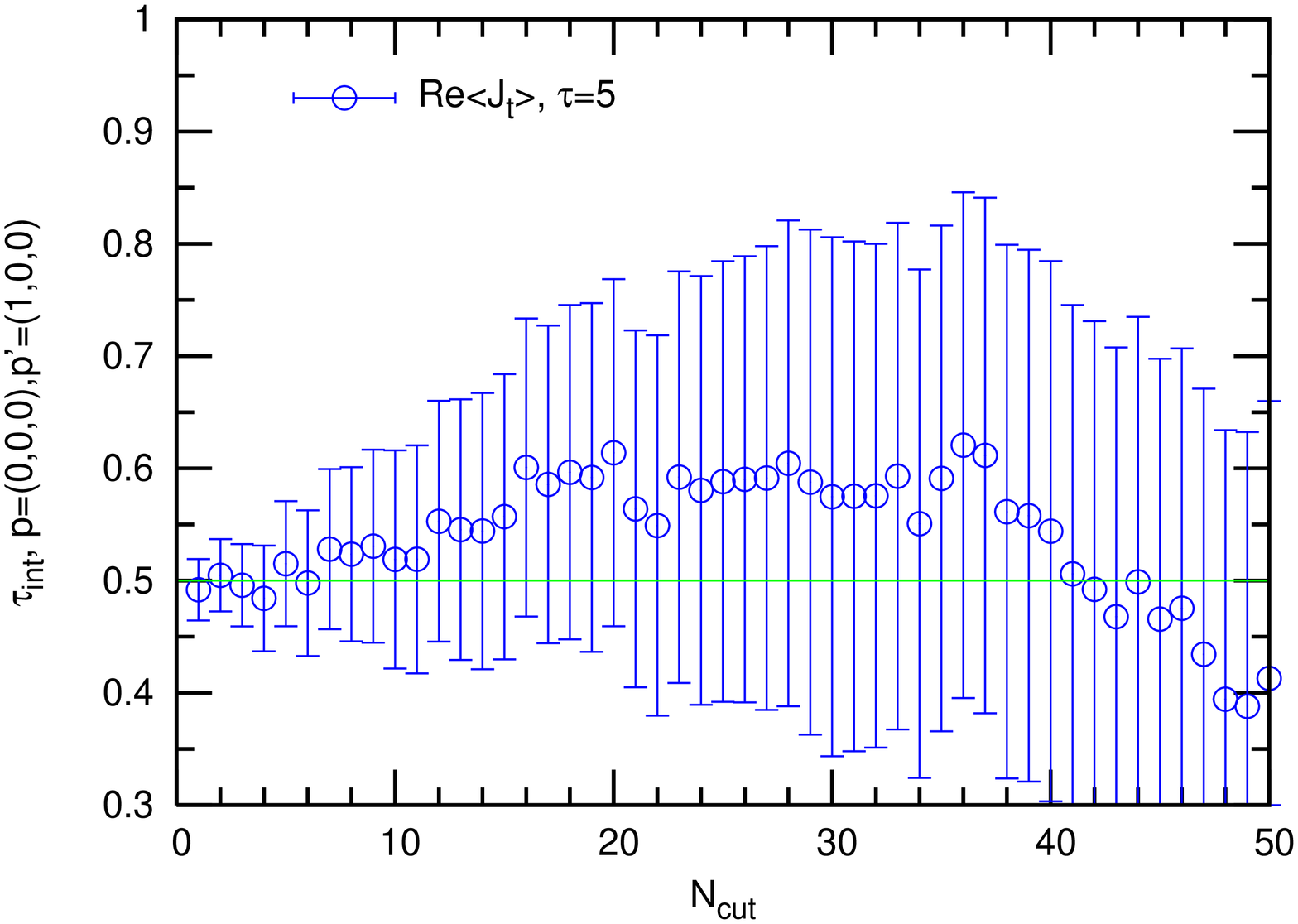}
\label{fig:int_threept}
}
\caption{Representative integrated autocorrelation time for two-point and three-point correlation functions. For completely decorrelated measurements, the integrated autocorrelation time is normalized to 1/2.}
\end{figure}

We calculated $\tau_{\rm int}$ in our measurements by treating each source as an individual measurement.  Figure~\ref{fig:int_twopt} shows the result for a zero-momentum projected two-point correlation function at a time separation $t=5$. The horizontal axis is the cut in the summation for the integrated autocorrelation time in terms of the number of measurements, which is 8 per lattice. One can see that $\tau_{\rm int}$ reaches a plateau at around 0.7, meaning the measurements are already de-correlated for every other sources. A similar result for the three-point correlation function is shown in Figure~\ref{fig:int_threept}, in which $\tau_{\rm int}$ is very close to 1/2, indicating that the measurements from adjacent sources are independent. These results are consistent with the binning study as discussed previously.

\subsection{Correlations Among Different Momentum Transfer}
\label{sec:corr_q}

\begin{figure}[t]
\centering
\subfigure[]{
  \includegraphics[width=0.45\textwidth]{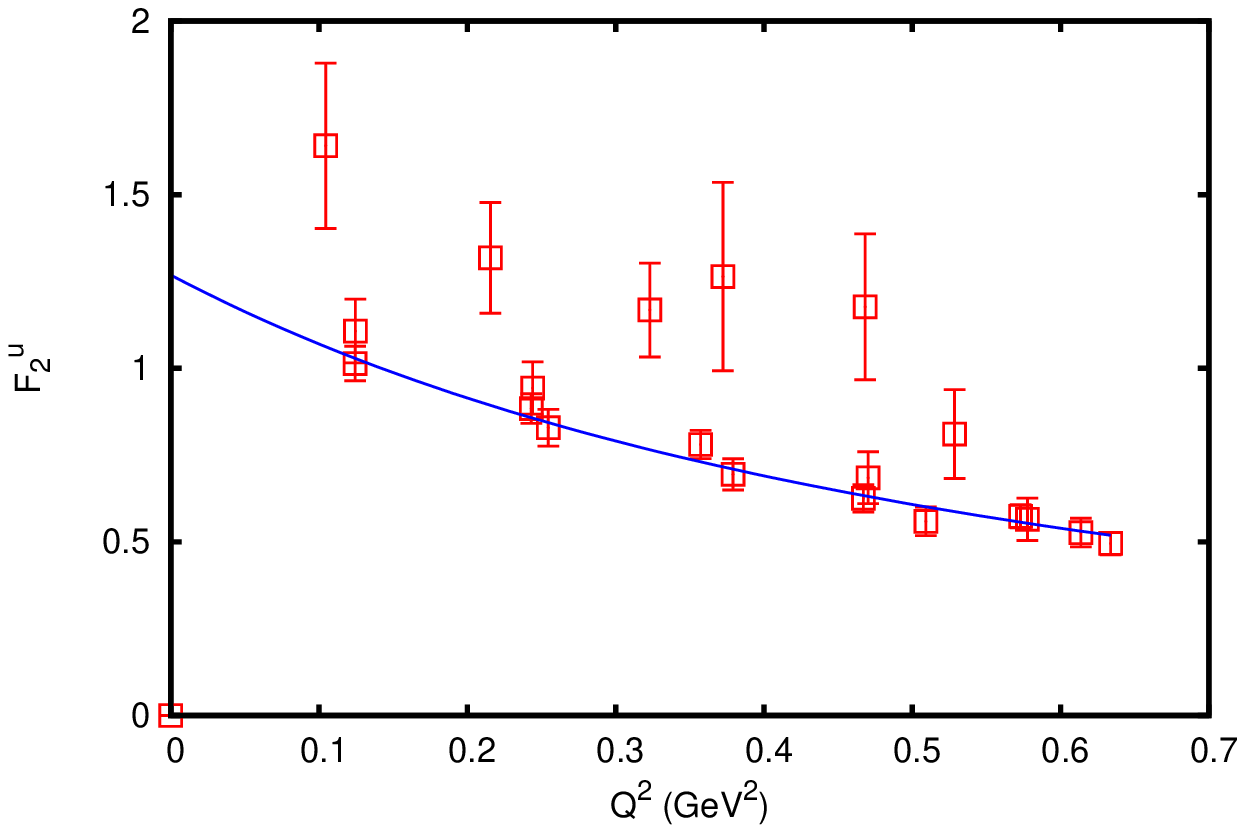}
  \label{fig:outlier_before}
}
\subfigure[]{
  \includegraphics[width=0.45\textwidth]{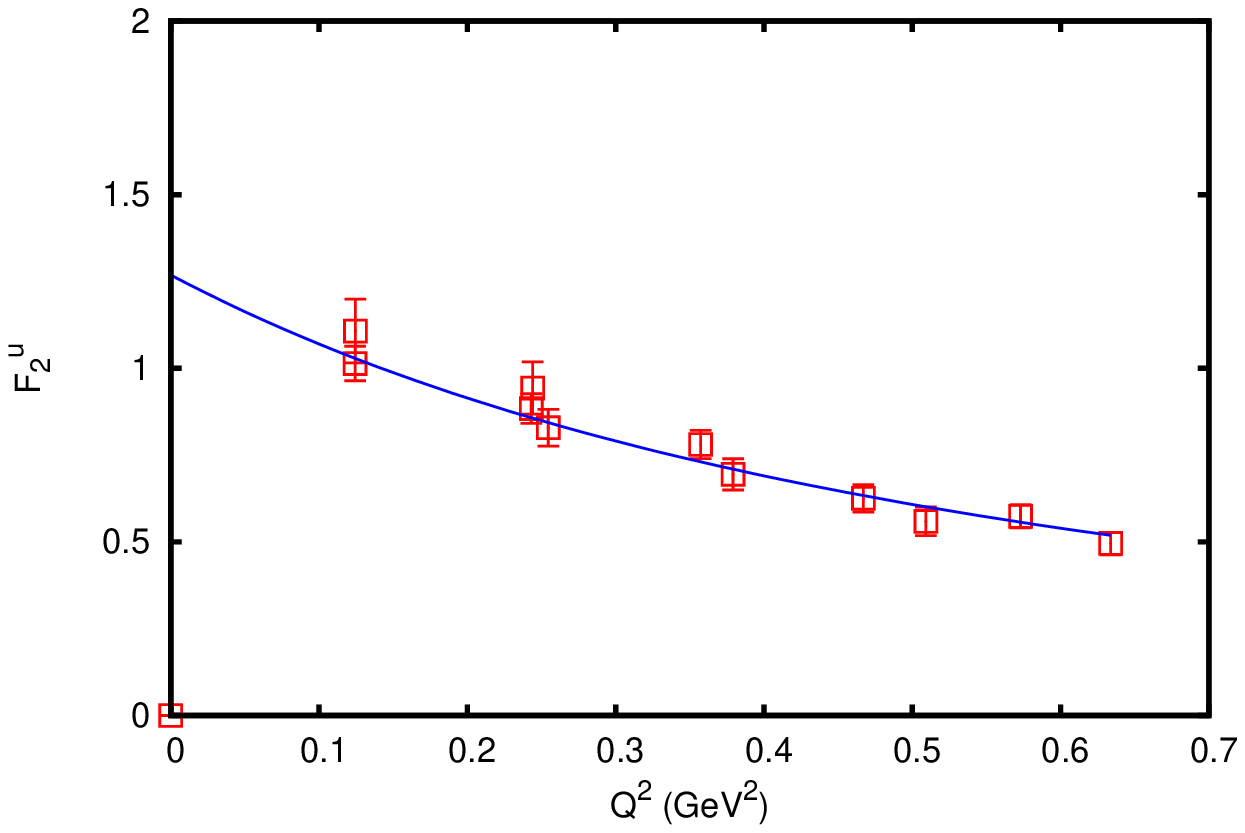} 
  \label{fig:outlier_after}
}
\caption{Example of overdetermined analysis for the $F_2$ form factor as a function of $Q^2$ including large boosts of source and sink. The left panel includes the source momentum components (-2,0,0) or larger, and sink momentum (-1,0,0). The right panel has had these data points removed.}
\label{fig:F2u_28}
\end{figure}

To maximize the hadron structure information determined from a given set of lattice  configurations, we use the overdetermined analysis introduced in Ref.~\cite{Schroers:2003mf} to simultaneously extract a specified set of generalized form factors from as many different combinations of twist-two operators and source-sink momenta as possible. One potential liability of this approach is the admixture of noisy measurements arising from source and sink momenta that are sufficiently high that the data are subject to large statistical errors.  If such data are obviously consistent with the accurate data, they do not affect chi-squared fits, and there is no significant bias in removing them from the analysis after the fact.  However, frequently the offending data gives the superficial appearance of being statistically inconsistent with the accurate data, if one ignores correlations. 
A typical example is the measurement of  $F_2^u$ on a  $28^3\times64$ lattice at $m_\pi \sim$ 350 MeV, as shown in Figure~\ref{fig:outlier_before}, where it appears that 6 out of 21 data points lie significantly above the reference curve. 

The question is whether 6 independent measurements lie several $\sigma$ off the
fit to the accurate data, which would be a highly statistically significant discrepancy, or if the data are highly correlated so that only one or two degrees of freedom have fluctuated randomly. Hence, we calculated the correlation matrix 
\begin{equation}
r_{ij} = C_{ij}/\sqrt{C_{ii} C_{jj}},
\end{equation} 
where the covariance matrix is defined as 
\begin{equation}
C_{ij} = (N - 1)\sum_{n=0}^{N-1} (F_i^{(n)} - \overline{F_i})(F_j^{(n)} - \overline{F_j}).
\end{equation}
 $F_i^{(n)}$ is the $n^{th}$ 
 jackknife sample of the $i^{th}$ momentum (hence the factor $N-1$), and $\overline{F_i}$ is the ensemble average of the $i^{th}$ momentum transfer. 
The resulting correlation matrix for the six outliers is 
\begin{equation}
r = \left (
\begin{array}{llllll}
1          & 0.822855 & 0.654945 & 0.639426 & 0.552542 & 0.520037 \\
0.822855  & 1  &      0.874902 & 0.649138 & 0.712735 & 0.742609 \\
0.654945  & 0.874902 & 1 &   0.562507 & 0.697565 & 0.701146 \\ 
0.639426  & 0.649138 & 0.562507 & 1 &  0.818721 & 0.408017 \\ 
0.552542  & 0.712735 & 0.697565 & 0.818721 & 1 & 0.584403 \\ 
0.520037  & 0.742609 & 0.701146 & 0.408017 & 0.584403 & 1 \\
\end{array}
\right )\, ,
\end{equation}
which shows very strong correlations between the data. 

To quantify the correlation, we calculated the $\chi^2$ of the outliers relative to a reference curve, which involves two steps.  First, we defined the reference curve by performing a correlated $\chi^2$ dipole fit to all  the data points (except for $Q^2=0$), including the six ``outliers''. Second, the $\chi^2$ of offending data points was calculated by
\begin{equation}
\chi^2 = N \sum_{n =0}^{N-1} (F_i^{(n)} - \hat{F_i})C^{-1}_{ij}(F_j^{(n)} - \hat{F_j}),
\end{equation}
where $\hat{F_i}$ is the expected form factor result at $i^{th}$ momentum from the fit, and $N$ is the number of outliers. If $\chi^2$ is close to 1, then the data points are consistent with the fit curve, and hence, the rest of the data points. In this example, the $\chi^2/N$ is found to be $1.9\pm1.1$, indicating that the deviation is not statistically significant. In contrast, a naive visual analysis of the curve ignoring correlations would lead to the erroneous conclusion that uncorrelated points are away from the curve by two $\sigma$, in which case  the $\chi^2/N$ would be 4.  This justifies the systematic exclusion of large source and sink momenta that give rise to noisy outliers, the result of which is shown in Figure~\ref{fig:outlier_after}.

\section{Control of Lattice Artifacts}
\label{sec:artifacts}

\subsection{Finite Volume Effects}
\label{sec:finiteV}
As we go to lighter pion masses while holding the lattice volume fixed, the effects of finite volume become more and more important. While it is widely accepted that $m_\pi L\geq 4 $ is necessary to avoid sizable finite volume effects, there are still controversies on how big the volume should be to have a good control over finite volume effects for nucleon physics. Here we do not attempt to address this question from a theoretical point of view. Rather, we present numerical evidence from our mixed-action calculations with two different volumes at a pion mass of roughly 350 MeV, to estimate how large the finite volume effects might be with this particular action and chosen lattice parameters.  

\begin{figure}[t]
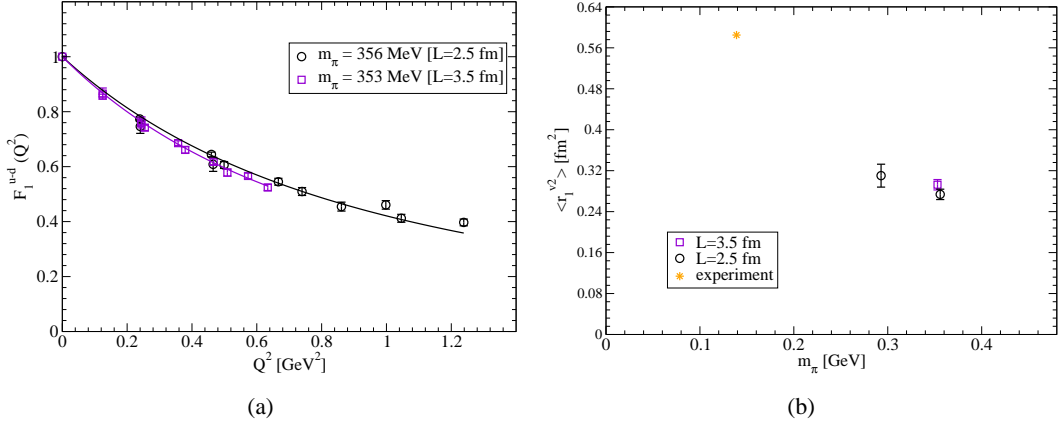

\centering
\subfigure[]{
\includegraphics[width=0.45\textwidth]{figs/F1umd_un_p01.eps}
\label{fig:FV-F1}
}
\subfigure[]{
\includegraphics[width=0.45\textwidth]{figs/radius.eps}
\label{fig:FV-r1}
}
\caption{(a) Isovector Dirac form factor from two different volumes. (b) Isovector Dirac radius.  }
\end{figure}

In Figure~\ref{fig:FV-F1}, we compare the results for the isovector Dirac form factor from the $20^3\times64$ and $28^3\times64$ ensembles, corresponding to physical volumes of roughly $(2.5 \mathrm{fm})^3$ and $(3.5 \mathrm{fm})^3$, respectively. The solid curves are dipole fits to the lattice data with $Q^2 \leq 0.4$ GeV$^2$. The fit parameters agree within errors, showing that there are no statistically significant differences between these two volumes. Since the slope of the form factor at $Q^2=0$ gives the Dirac radius of the nucleon, one can infer that the results for the Dirac radius obtained from these two volumes do not show significant finite volume effects. We present the comparison of the Dirac radii in Figure~\ref{fig:FV-r1}, where we also include the result at the pion mass of 293 MeV. And the star is the phenomenological value as obtained in~\cite{Mergell:1995bf}.

\begin{table}[t]
\centering
\begin{tabular}{|c|c|c|c|}
\hline
& $20^3\times64$ ($L\approx2.5$ fm) & $28^3\times64$ ($L\approx3.5$ fm) & $\Delta_V$ \\
\hline
$\langle {r_1^{v}}^2  \rangle$ [fm$^2$]& 0.274(10) & 0.290(10) & -0.016(14)\\
\hline
$g_A$ & 1.161(14) & 1.153(14) & 0.008(20) \\
\hline
\end{tabular}

\caption{\label{tab:FV}Comparison of results for $\langle {r_1^{v}}^2 \rangle$ and $g_A$ from two different lattice volumes. $\Delta_V$ is the difference between the $L=2.5$ fm result and the $L=3.5$ fm result. The error on $\Delta_V$ is calculated by adding errors from the two volumes in quadrature.}
\end{table}

To estimate the finite volume effects quantitatively, we calculate the differences, $\Delta_V$, from the small-volume (2.5 fm) and large-volume (3.5 fm) calculations for the isovector Dirac radius, $\langle {r_1^{v}}^2  \rangle$, and nucleon axial charge, $g_A$. Since these two ensembles are statistically independent, we simply calculate the errors on $\Delta_V$ by adding errors in quadrature. The results are given in Table~\ref{tab:FV}. We can see that the differences are statistically consistent with zero, suggesting that the finite volume effects are negligible compared to statistical errors.

\subsection{Comparison of Domain-Wall and Mixed Actions}
\begin{figure}[t]
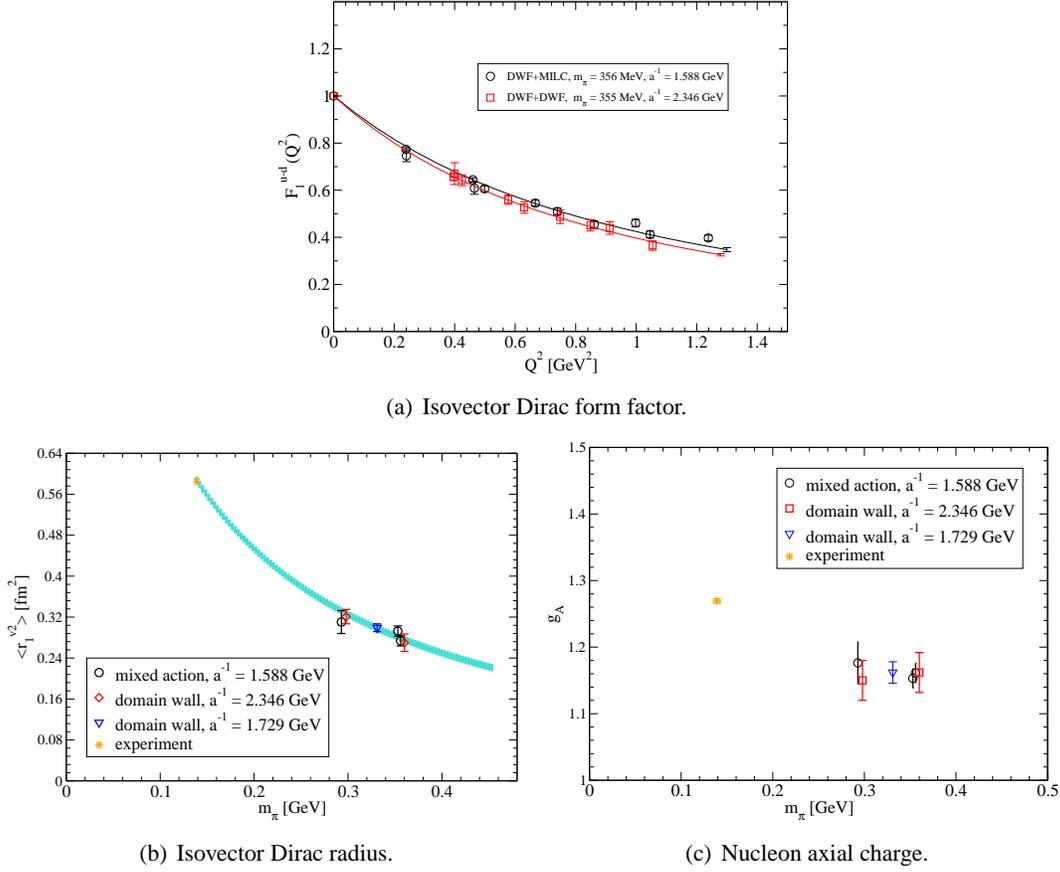

\centering
\subfigure[Isovector Dirac form factor.]{
\includegraphics[width=0.45\textwidth]{figs/comp_MILC_DWF_350.eps}
\label{fig:comp_ff}
}\\
\subfigure[Isovector Dirac radius.]{
\includegraphics[width=0.45\textwidth]{figs/radius_w_DW.eps}
\label{fig:comp_r1}
}
\subfigure[Nucleon axial charge.]{
\includegraphics[width=0.45\textwidth]{figs/gA_w_DW.eps}
\label{fig:comp_gA}
}
\caption{Comparison of some physical results from the mixed-action and domain-wall calculations. In (b) and (c), the $a^{-1} = 2.346$ GeV domain wall results are slightly shifted to the right for clarity. }
\end{figure}

\label{sec:discretization}
Since domain wall fermion formulation is automatically $O(a)$ improved, we expect the discretization error in the full domain wall calculation to be small. In our full domain wall calculations of the nucleon form factors, this is found to be the case, as described in Ref.~\cite{Syritsyn1}. To assess the discretization error of our mixed-action calculations, here we compare some of the nucleon structure results from both actions as available at the time of the lattice conference. For latest domain wall results with improved statistics, see Ref.~\cite{Syritsyn1}.

In Figure~\ref{fig:comp_ff} we show the isovector Dirac form factor from the fine ($a^{-1} \approx 2.346$ GeV) domain wall and the coarse ($a^{-1} \approx 1.588$ GeV) mixed-action calculations, both with a pion mass of roughly 350 MeV. Over the whole range of $Q^2$ available to us, both actions give statistically consistent results. We note that the lattice scale of the fine domain wall ensembles used here is a crude estimate obtained in Ref.~\cite{Syritsyn1}. Using the Sommer parameter, the lattice scale for the fine domain wall enembles was found to be $a^{-1}=2.42(4)$ GeV~\cite{Scholz:2008uv}. If the latter were used, the $Q^2$ values would shift to the right, and the pion mass would become larger. The combined effect would shift the domain wall results upward to be even more consistent with the mixed-action data,  which is an indication that the discretization error for the mixed-action calculation is small.

Figure~\ref{fig:comp_r1}  shows the isovector Dirac radius from the coarse mixed-action calculation and domain wall calculations at two different lattice spacings. The curve is a one parameter ($B_{10}^r$) fit to the three mixed action data points using the $O(\epsilon^3)$ small-scale-expansion (SSE) chiral formula as given in~\cite{Bernard:1998gv}, with the low energy constants fixed to phenomenological values. One can see that all the data points fall on the same curve, showing that the mixed action results qualitatively agree with the domain wall data. The same holds true for the nucleon axial charge, as shown in Figure~\ref{fig:comp_gA}, where all the data show little pion mass dependence, and all lie on one horizontal line which is a few percent below the experimental result. 

Since in both the mixed-action and the fine domain wall calculations, the lightest two pion masses are comparable ($\sim 300$ and 350 MeV), we are able to do a more quantatitive comparison. Table~\ref{tab:comp_MA_DW} gives the differences between the mixed-action (MA) and domain-wall (DW) results, $\Delta_a$, for $\langle {r_1^{v}}^2 \rangle$  and $g_A$, at these two pion masses. Once again, the differences are found to be consistent with zero within errors.  

%
%
\begin{table}[t]
\centering
\begin{tabular}{|c|c|c||c|c|}
\hline
& \multicolumn{2}{|c||}{$\langle {r_1^{v}}^2 \rangle$ [fm$^2$]} & \multicolumn{2}{|c|}{$g_A$} \\
\hline
&  $m_\pi$ $\approx 300$ MeV &  $m_\pi$ $\approx 350$ MeV & $m_\pi$ $\approx 300$ MeV  &  $m_\pi$ $\approx 350$ MeV\\
\hline
MA & 0.311(22) & 0.274(10) & 1.176(32) & 1.161(14) \\
\hline
DW & 0.321(14) & 0.270(17) & 1.150(30) & 1.162(30) \\
\hline
$\Delta_a$ & - 0.010(26) &  0.004(20) &  0.026(44) & -0.001(33) \\
\hline
\end{tabular}
\caption{Estimates of differences between mixed-action and domain wall calculations for $\langle {r_1^{v}}^2 \rangle$  and $g_A$. $\Delta_a$ is the difference between the mixed-action and domain-wall action results. The error on $\Delta_a$ is calculated by adding the errors from the two actions in quadrature.}
\label{tab:comp_MA_DW}

\end{table}

\section{Select Physical Results}
\label{sec:results}

\subsection{Electromagnetic Form Factors}
\label{sec:EMff}
\begin{figure}[t]
  \centering
  \includegraphics[width=0.5\textwidth,angle=-90]{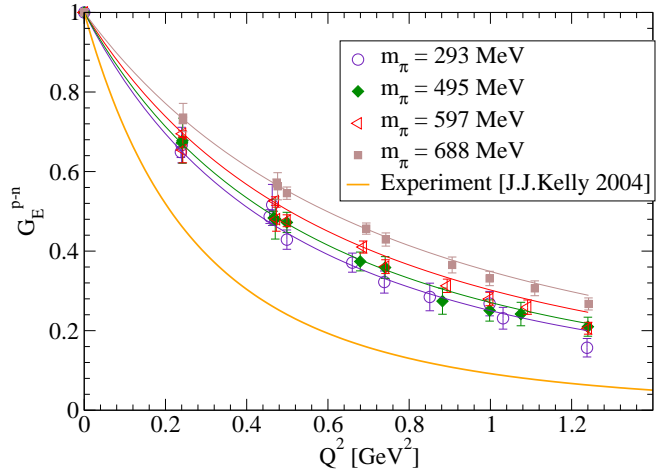}
  \caption{Isovector electric form factor from the mixed-action lattice calculations along with the Kelly parametrization~\cite{Kelly:2004hm} of the experimental data.}
  \label{fig:Ge}
\end{figure}

Complementary to the results for the nucleon electromagnectic form factors mentioned in previous discussions, in Figure~\ref{fig:Ge} we show the isovector electric form factors from several masses, including results from previous calculations~\cite{Edwards:2006qx} at pion masses of 495 MeV, 597 MeV and 688 MeV. To avoid duplication, we do not show results with the pion mass of about 350 MeV, which was already discussed in Section~\ref{sec:finiteV}. The curves through the lattice data are one-parameter dipole fits. The bottom solid curve is Kelly's parametrization of the experimental results~\cite{Kelly:2004hm}. While in previous sections, we have seen that the chirally extrapolated lattice result for the Dirac radius is compatible with the experiment, here we see that over the range of momentum transfer up to $Q^2 \leq 1.2$ GeV$^2$, the lattice results show a monotonic decrease towards the experimental curve as the pion mass gets smaller. To have a direct comparison with the experiment for the dependence on the momentum transfer, we need to perform a chiral extrapolation with $m_\pi$ and $Q^2$ dependences taken into account simultaneously~\cite{Bernard:1998gv}, which is still a work in progress.  

\subsection{Quark Angular Momentum Contribution to the Nucleon Spin}
\begin{figure}[t]
  \begin{tabular}{cc}
  \hspace{-.3cm}  \includegraphics[width=0.48\textwidth]{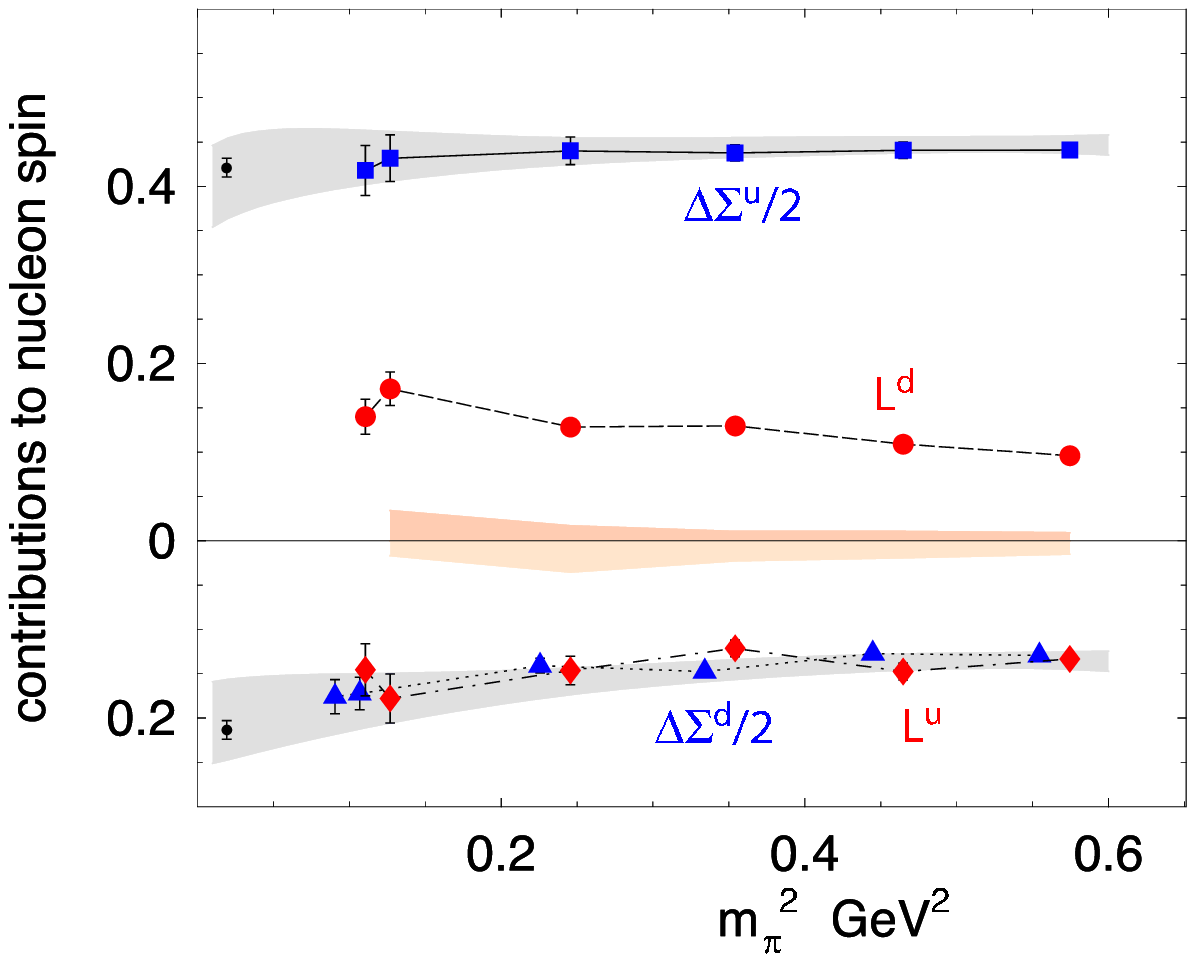} &
   \raisebox{1.5 cm}{ \includegraphics[width=0.48\textwidth]{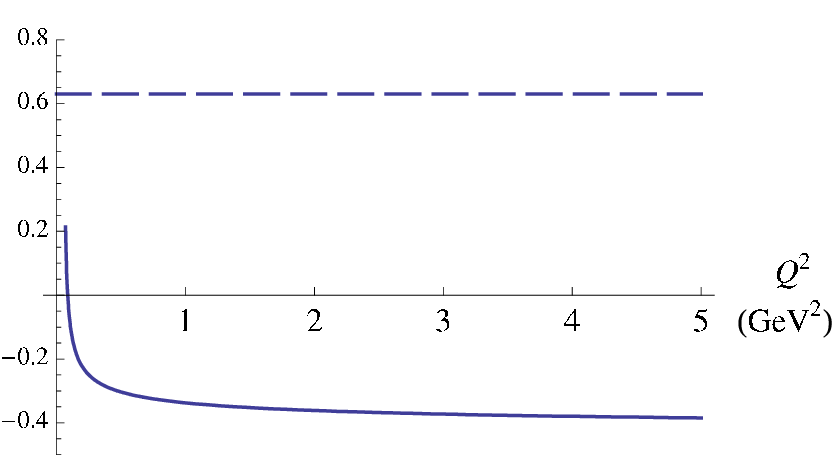}}\\
  \end{tabular}
  \caption{\label{spin} The left panel shows the quark spin and orbital angular momentum contributions to the nucleon spin from ref.\cite{Hagler:2007xi}. The right panel show evolution of the quark angular momentum with respect to the scale $Q^2$.}
\end{figure}

The origin of the nucleon spin is a forefront question in contemporary experiment and theory. Figure \ref{spin} shows our recent lattice calculation~\cite{Hagler:2007xi} of the quark spin and orbital contributions to the nucleon spin, which makes a strong case that the  sign of the spin contribution of a given flavor, $\Delta \Sigma$ and the sign of the  orbital contribution $L$ for the same flavor are opposite at the scale 4 GeV$^2$ at which the lattice calculation was renormalized. 

Since this behavior is strikingly different from simple familiar models, it is interesting to consider its origin and significance.  In a mean field model, in which a quark satisfies the Dirac equation in a central potential, for a nucleon with spin projection in the positive $z$ direction, the upper component is an $S$-state with spin up and the lower component is a $P$-state with orbital angular momentum projection +1 and spin down. Thus, generically, the spin and orbital angular momentum are aligned in mean field theory.  Clearly, the reason we solve QCD from first principles is to go beyond models, and there is no reason that the nucleon should necessarily be consistent with mean field theory.  However, stimulated by a recent paper by Thomas \cite{Thomas:2008ga}, it is interesting to note that the sign of $L^{u-d}$ is strongly scale dependent.  

We restrict our attention to the flavor non-singlet sector, for which there are no disconnected quark diagrams, and no mixing with gluons, and consider the evolution of orbital and spin contributions shown in the right panel of Figure~ \ref{spin}.  The spin contribution $\Delta \Sigma^{u-d}$ is large and is conserved under QCD evolution as shown by the dashed line. However, at one-loop level the total angular momentum has the simple evolution given by its anomalous dimension
$$
L^{u-d}(t) + \frac{\Delta \Sigma^{u-d}}{2}  = 
\left( \frac{t}{t_0} \right)^{\frac{32}{81}}
\left( L^{u-d}(t_0) + \frac{\Delta \Sigma^{u-d}}{2} \right),
$$
where $ t = \ln ( \frac{Q^2}{\Lambda^2_{QCD}} ) $.
  Because the largest contribution to the total angular momentum is the spin, which cannot evolve, the full change under evolution must arise from the orbital contribution $L^{u-d}$, which must therefore vary significantly with the scale. As shown by the solid line in the right panel, the change required in $L^{u-d}$ is so substantial that it changes sign when the scale becomes low enough. Although surely one loop evolution is not quantitatively reliable and evolution below 1 GeV is suspect, it is clear that even the relative sign between spin and orbital contributions is scale dependent, so there is no reason our lattice calculation at 4 GeV$^2$ need to be consistent with simple quark models or mean field arguments.

%


\section{Conclusions}
\label{sec:conclusions}
We have investigated several aspects of the nucleon form factor calculations in an attempt to address some issues which are essential for achieving high precision calculation in the chiral regime. In particular, as the numerical cost increases substantially as our calculations move to lighter pion masses, we employed the coherent sink technique in the calculation of the backward quark propagator to reduce the computational cost. To obtain reliable statistical errors, we also studied the autocorrelations in the measurements and found no strong correlations between measurements from adjacent sources. Our results at a pion mass of 350 MeV from two lattice volumes, (2.5fm)$^3$ and (3.5fm)$^3$, show no significant differences, suggesting negligible finite volume effects. We also compared the mixed-action results with the full domain wall calculations at a finer lattice spacing, and found that the discretization errors from using the mixed action is quite small. 

\section*{Acknowledgements}

This work was supported in part by U.S. DOE Contract No. DE-AC05-06OR23177 under which JSA operates Jefferson Laboratory, by the DOE Office of Nuclear Physics under grants 
DE-FG02-94ER40818, 
DE-FG02-04ER41302, 
DE-FG02-96ER40965, 
by the DFG (Forschergruppe Gitter-Hadronen-Ph\"anomenologie), and  the EU Integrated Infrastructure Initiative Hadron Physics (I3HP) under contract  RII3-CT-2004-506078. 
W.S. acknowledges support by the National Science Council of Taiwan under grants  NSC96-2112-M002-020-MY3 and NSC96-2811-M002-026, 
K.O. acknowledges support from the  Jeffress  Memorial Trust grant J-813,
Ph. H. and B. M. acknowledge support by the Emmy-Noether program of the DFG, the Excellence Cluster Universe at
the TU Munich, and  Ph. H., M.P. and W. S. acknowledge support by the A.v. Humboldt-foundation through the Feodor-Lynen program.  This research used resources under the INCITE and ESP programs of the Argonne Leadership Computing Facility at Argonne National Laboratory, which is supported by the Office of Science of the U.S. Department of Energy under contract DE-AC02-06CH11357, resources provided by the DOE through the USQCD project at Jefferson Lab and through its support of the MIT Blue Gene/L under grant DE-FG02-05ER25681,  resources provided by the William and Mary Cyclades Cluster, and resources provided by the New Mexico Computing Applications Center (NMCAC) on Encanto.  We are indebted to members of the MILC, RBC, and UKQCD Collaborations for providing the dynamical quark configurations that made our full QCD calculations possible.

\bibliography{reference}

\bibliographystyle{JHEP-2} 

\end{document}